# Voxel-Level Brain States Prediction Using Swin Transformer

Yifei Sun, Daniel Chahine, Qinghao Wen, Tianming Liu, Xiang Li, Yixuan Yuan, Fernando Calamante, Jinglei Lv.

***Abstract*—Understanding brain dynamics is important for neuroscience and mental health. Functional magnetic resonance imaging (fMRI) enables the measurement of neural activities through blood-oxygen-level-dependent (BOLD) signals, which represent brain states. In this study, we aim to predict future human resting brain states with fMRI. Due to the 3D voxel-wise spatial organization and temporal dependencies of the fMRI data, we propose a novel architecture which employs a 4D Shifted Window (Swin) Transformer as encoder to efficiently learn spatio-temporal information and a convolutional decoder to enable brain state prediction at the same spatial and temporal resolution as the input fMRI data. We used 100 unrelated subjects from the Human Connectome Project (HCP) for model training and testing. Our novel model has shown high accuracy when predicting 7.2s resting-state brain activities based on the prior 23.04s fMRI time series. The predicted brain states highly resemble BOLD contrast and dynamics. This work shows promising evidence that the spatiotemporal organization of the human brain can be learned by a Swin Transformer model, at high resolution, which provides a potential for reducing the fMRI scan time and the development of brain-computer interfaces in the future.***

## I. INTRODUCTION

The human brain operates as a highly complex and dynamic system [1, 2], with neural activity continuously evolving across space and time [3], during tasks as well as during resting states [4]. Understanding these brain dynamics is essential for uncovering the neural basis of various functions, such as cognition, emotion, and memory [5]. Moreover, capturing how brain states unfold can provide insights into brain disorders [6-8] and potentially guide diagnostic and therapeutic strategies.

Functional magnetic resonance imaging (fMRI) offers a non-invasive means of indirectly observing brain activities by measuring the blood-oxygen-level-dependent (BOLD) signals [9]. Despite various analysis methods developed for fMRI data [10-14], a fundamental challenge or gap remains: how do brain states arise at the resting state, and can future brain state be accurately predicted from previous activity? Solving this problem could enhance our understanding of intrinsic and dynamic brain processes. More practically, it can help reduce the fMRI scanning time for participants, especially for patients who cannot stay in the scanner for a long time. Furthermore, by learning the human brain mechanism, brain states prediction has the potential to be applied to the brain-computer interface (BCI) [15] or assist the understanding of the abnormal brain dynamics of brain disorders, such as epilepsy and Parkinson's disease.

In recent years, deep learning models, particularly transformer architectures [16], have revolutionized sequence modeling tasks. With self-attention [16], transformer models were originally applied in natural language processing because of their extraordinary ability in capturing long-range dependencies in sequential data. This capability has enabled breakthroughs in artificial intelligence, such as ChatGPT [17], which showcases the power of transformers in understanding and generating human language. Beyond natural language processing, transformers have demonstrated strong potential in modeling other types of sequential data, including time series from fMRI [18-20], where capturing temporal dependencies is critical for understanding dynamic brain processes.

While graph networks [13, 14] and transformers [19, 21]

J. Lv is supported by Brain and Mind Centre Research Development Grant, USYD-Fudan Brain and Intelligence Science Alliance Flagship Research Program, Moyira Elizabeth Vine Fund for Research into Schizophrenia Program and ARC Discovery Project (DP240102161). D. Chahine is supported by TOM YIM scholarship, USYD Vacation Research Internship Program, Engineering Accelerator Scholarship, and Faculty of Engineering, University of Sydney.

Corresponding author: J. Lv.
Y. Sun is with the School of Biomedical Engineering and Brain and Mind Center at the University of Sydney, Sydney, NSW 2050 Australia (e-mail: yifei.sun@sydney.edu.au).
D. Chahine is with the School of Biomedical Engineering at the University of Technology Sydney, Sydney, NSW 2007 Australia (e-mail: daniel.chahine@student.uts.edu.au).
Q. Wen is with the School of Aerospace, Mechanical and Mechatronic Engineering, the University of Sydney, Sydney, NSW 2050 Australia (e-mail: qwen2355@uni.sydney.edu.au).
T. Liu is with the School of Computing, University of Georgia, Athens, GA 30602, USA (e-mail: tliu@uga.edu).
X. Li is with the Department of Radiology, Massachusetts General Hospital and Harvard Medical School, Boston, MA 02114, USA (e-mail: xli60@mgh.harvard.edu).
Y. Yuan is with the Department of Electronic Engineering, Chinese University of Hong Kong, Hong Kong SAR 999077, China (e-mail: yxyuan@ee.cuhk.edu.hk).
F. Calamante is with the School of Biomedical Engineering, Brain and Mind Center, and Sydney Imaging, at the University of Sydney, Sydney, NSW 2050 Australia (e-mail: fernando.calamante@sydney.edu.au).
J. Lv, is with the School of Biomedical Engineering and Brain and Mind Center at the University of Sydney, Sydney, NSW 2050 Australia (e-mail: jinglei.lv@sydney.edu.au).

have been adopted to study brain dynamics, most of which simplified brain states to regional level, overlooking the rich spatial details available at the voxel level. Particularly, voxel-wise modelling requires handling 4D fMRI data composed of 3D voxels with their time series, which dramatically increases data dimensionality. In our previous work [21], we leveraged a basic transformer model for brain state prediction based on 379 brain regions, showing the potential of transformers in such tasks. However, standard transformer architectures are not optimized for such high-resolution spatiotemporal data, limiting their scalability and effectiveness in voxel-level modeling.

The Shifted Window (Swin) Transformer [22], a variant of the Vision Transformer [23], addresses these limitations through a hierarchical attention mechanism and window-based processing. This design enables efficient modelling of both local and global dependencies in high-dimensional data. Recent studies proposed a Swin 4D fMRI Transformer, called SwiFT [18], and showed its impressive ability in age, sex, and cognitive trait prediction, demonstrating its suitability for learning complex spatiotemporal information from neuroimaging data.

Inspired by these advances, we propose a modified SwiFT architecture to tackle the challenge of voxel-wise brain state prediction. In our approach, each 3D fMRI volume at specific time point is treated as a brain state, and the temporal evolvement of brain activity is modeled as a sequence of such volumes. By feeding a sequence of preceding fMRI volumes into the adapted Swin Transformer model, we aim to predict a subsequent set of brain states, and to achieve this at the whole brain voxel level. This formulation enables the model to learn fine-grained spatiotemporal patterns across both space and time. Our approach not only extends the application of Swin transformers in fMRI analysis but also introduces a new framework for forecasting fine-grained brain activity, potentially contributing to more efficient scanning protocols and deeper insights into resting-state brain dynamics.

## II. METHODS

### A. Data Preparation

We used 100 unrelated young adults' resting state fMRI (rs-fMRI) data from the Human Connectome Project (HCP) [24]. These subjects have an age range from 22 to 36, with 54 females and 46 males. For each subject, we used two minimally preprocessed [25] 3 Tesla (3T) scans from session 1 (REST1) with the left-to-right and right-to-left phase encoding directions, respectively. Each rs-fMRI scan has 1200 time points in the volume space with an isotropic spatial resolution of 2mm and a temporal resolution of 0.72s.

We carried out several preprocessing steps to further clean the data, including spatial smoothing using a Gaussian filter with a full width at half maximum (FWHM) of 6mm and bandpass filtering, retaining signals between 0.01 and 0.1 Hz. Finally, time series at each voxel were z-score transformed to have zero temporal mean and unit standard deviation, so that they have similar scale, which could potentially help the model convergence.

For each subject, we also have their brain mask in the same space. These masks are used to restrict the evaluation of the model predictions solely in the brain area, i.e. disregarding the background.

The 2mm voxel-level rs-fMRI data contains rich information, however, the high dimensionality poses challenges for model training and testing. Considering our computing capacity, we down-sampled the fMRI data by halving the spatial dimensions using cubic interpolation but preserving the original temporal resolution. Specifically, the original fMRI data dimension of *91×109×91×1200*, was reduced to *46×54×46×1200*.

To ensure compatibility with the SwiFT model and enable efficient batch processing, all fMRI volumes were spatially uniformed to a consistent shape of *48×48×48* voxels. For dimensions exceeding the target size, central cropping was applied; for smaller dimensions, symmetric zero-padding was utilized. The temporal dimension, i.e. number of time points, was preserved throughout the process.

### B. Model Architecture

In our new architecture, the SwiFT [18] model was used as a encoder aiming to learn the spatiotemporal patterns in the input fMRI time series. To enable the model to output 3D volumes, we added a convolutional decoder containing three 3D transposed convolutional layers. This decoder upsampled the data to output future brain states with the same spatial resolution as the input. The overview of the model architecture is depicted in Fig. 1.

As shown in Fig. 1, the SwiFT encoder takes the 4D fMRI time series with the shape of *W×H×D×T×1* as input, where *W, H, and D* are the spatial dimensions and *T* stands for the number of consecutive brain states in the input time series with one channel. The encoder first split the input image into small patches of the shape *p×p×p×t* using patch partitioning. Then, we have the patch embedding and positional embedding to convert the $tp^3$ patches into E embedding dimensions (we used *E=36*), i.e. tokens. After that, the embedded tokens were passed into the first 4D Swin Transformer block with two layers with three heads. Starting from the second stage, the patch merging further reduces the dimensionality. In stage 2 and 3 Swin Transformer blocks, we used six layers with six heads and two layers with twelve heads, respectively. During our training and testing, our model will only get the preceding brain states as input and predict the future ones. This ensures the model will never tend to the future brain states for the prediction, so we omitted the use of look-ahead masking, which is mainly used to prevent the model from using future tokens during training and inference [16].

For the model working for the down-sampled rs-fMRI, the input brain volumes have the shape of *48×48×48×32×1*, equivalent to 23.04s fMRI data. We used a patch size of *3×3×3×4*. Both spatial and temporal window sizes were set to be four. The SwiFT encoder will hierarchically down-sample the input to *4×4×4×8×144*, while learning the spatiotemporal information.

The encoder output was then passed to the decoder for restoring the spatial dimensions and outputting the required number of brain states with the shape of $W \times H \times D \times T' \times 1$, where $T'$ is the number of predicted brain states. Here, we use the $T' = 10$ (7.2s), corresponding to the 10 sequential brain states right after the input brain states. We designed two skip connections between the encoder and decoder, as inspired by U-Net structure [26]. We first combined the temporal and channel dimensions, after which we treated the channels in the decoder as time points so that multichannel predictions can be regarded as multiple brain states. This enables the model to predict 10 brain states all at once. In this decoder, we utilized three 3D transpose convolutional layers for up-sampling. Two Skip connections passed the spatiotemporal maps from encoder to decoder. After each concatenation, we employed two 3D convolutional layers to help fuse channels. After the last up-sampling 3D transposed convolutional layer, we used a final 3D convolutional layer with the kernel size of one to tailor the output channel size to match the expected output length.

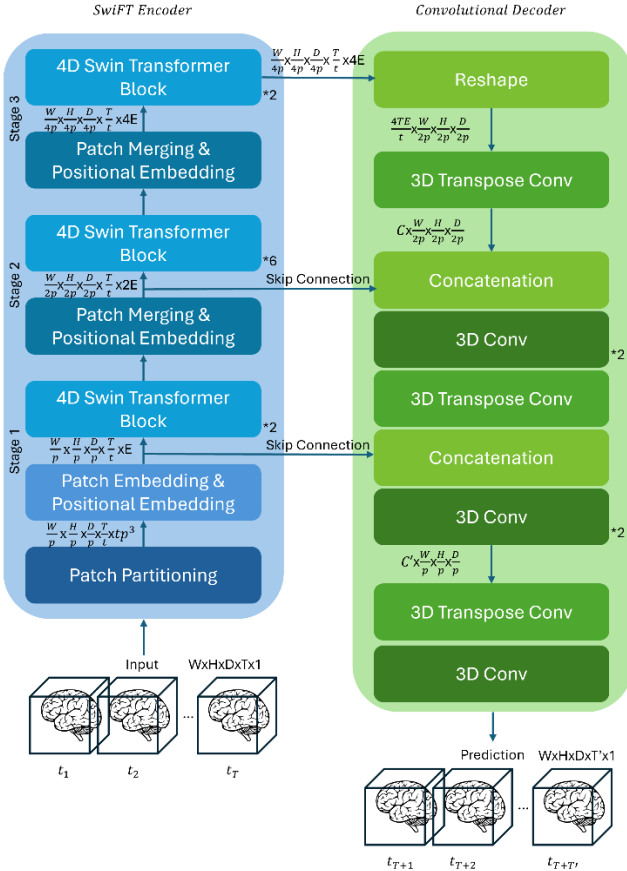

Fig. 1. The summary of our model architecture. Our model has an encoder-decoder structure with skip connections linking them. The encoder is the SwiFT and the decoder contains transpose convolutional and convolutional layers to up-sample the data to generate predictions. *W, H,* and *D* are the spatial dimensions of the input fMRI volumes. *T* is the number of consecutive brain states in the input. *T'* is the number of output brain states. *E* denotes the embedding dimensions, *p* is the initial spatial patch size, and *t* is the initial temporal patch size. *C* and *C'* are the number of channels.

### C. Training

Among 100 subjects, we randomly selected 80 subjects for training and the other 20 subjects for testing. For each fMRI scan, we used sliding window to generate the input and output samples. Each input sample contains 32 sequential brain states represented by 3D volumes, and the next 10 brain states are regarded as the ground truth brain states for evaluation.

We are only interested in the prediction of signals within the brain region, we used the masked mean squared error (MSE) for evaluation as defined in (1), which only considers the *n* brain voxels.

$$\text{Masked MSE} = \frac{1}{n}\sum_{i=1}^{n}(Y_i - \hat{Y}_i)^2 \quad (1)$$

Additionally, we evaluated the preservation of the contrast level in predicted brain states by the structural similarity index measure (SSIM) [27]. This metric measures the similarity between predicted and real brain states, by comparing local structural information within small patches. Specifically, we extended the conventional SSIM which is commonly applied to 2D images, to 3D space by employing a 3D Gaussian kernel as a sliding window to compute local statistics across the volumetric inputs. With padding, we calculated *m* patch centered at each voxel. Taking the ground truth patch *x* and predicted patch *y*, *SSIM(x, y)* (2) was calculated using local means ($\mu_x$ and $\mu_y$), variances ($\sigma_x$ and $\sigma_y$), and covariances ($\sigma_{xy}$) obtained via 3D convolution operations, and the resulting similarity maps were averaged across all *m* patches to produce a global SSIM score (3). Constants $c_1 = 0.01^2$ and $c_2 = 0.03^2$ in (2) are included to stabilize the division in regions of low contrast. These values are chosen as the default value for SSIM calculation not scaled by the pixel intensity range. The Gaussian window had a size of $7 \times 7 \times 7$ and was constructed with a standard deviation proportional to the window size.

In our task, the predicted brain states should not only be numerically close to the real brain states (i.e. with low MSE) but also preserve the anatomical structures and contrast (i.e. with high SSIM). In practice, we optimized the model to achieve higher SSIM in the prediction. With this aim, we define SSIM loss in (4), to enhance the preservation of contrast and textures in the predictions.

$$SSIM(x,y) = \frac{(2\mu_x\mu_y + c_1)(2\sigma_{xy} + c_2)}{(\mu_x^2 + \mu_y^2 + c_1)(\sigma_x^2 + \sigma_y^2 + c_2)} \quad (2)$$

$$SSIM = \frac{\sum_1^m SSIM(x_i, y_i)}{m} \quad (3)$$

$$SSIM\ Loss = 1 - SSIM \quad (4)$$

We added the SSIM loss as a penalty term to the model. The combined loss function we used is specified in (5). One factor $\alpha = 2$ was added to SSIM loss to make a similar initial scale as the masked MSE.

$$Loss = masked\ MSE + \alpha(1 - SSIM) \quad (5)$$

Our model was trained with an Adaptive Moment Estimation (Adam) optimizer, and the initial learning rate was set to be 1e-4. We monitored the training loss, masked MSE, and SSIM loss changes during the training of 24 epochs. This model was trained on our local server with NVIDIA A100 GPU cores and the CUDA version of 12.2.

## D. Prediction Performance

Model performance was evaluated using 20 testing subjects that were not seen during the training process. To quantitatively assess prediction accuracy, we computed the overall masked MSE and SSIM loss across all test samples and all predicted time points. These metrics reflect both voxel-wise intensity differences and structural preservation between predicted and ground truth brain states.

To determine whether the model captures temporal dependencies among input brain states rather than relying solely on static spatial patterns and the presence of single brain state, we conducted a control analysis by randomly shuffling the order of the 32 input time points. We then re-ran the prediction and compared the resulting MSE with that from the unshuffled input.

As we observed that adjacent time points in rs-fMRI are highly similar, we introduce the shifted MSE. This was defined by comparing the predicted brain state at time $t_i$ with the actual brain state at time $t_{i-1}$, rather than the true $t_i$. This analysis evaluates whether the model is replicating the adjacent brain state rather than learning meaningful transitions between brain states.

To further examine prediction dynamics over time, we evaluated the model's performance at each of the 10 output brain states individually. For each predicted state from $t_{T+1}$ to $t_{T+10}$, we computed the masked MSE across all test samples. This allowed us to characterize how prediction accuracy changes as the model forecasts further into the future. We also assessed the spatial distribution of voxel-wise prediction errors by visualizing the absolute difference for each predicted brain state at different anatomical planes.

## E. Qualitative Assessment

In voxel-level brain state predictions, preserving spatial features is critical. A low MSE alone can be insufficient to ensure meaningful predictions, as the model could achieve low error by generating smoothed outputs with limited contrast or predicting values close to the global mean. Therefore, we visualized random slices from predicted volumes and compared them to the corresponding ground truth brain states slices. This qualitative analysis allowed us to evaluate the preservation of spatial contrast, fine-grained anatomical detail, and dynamic changes across brain regions.

## F. Regional Prediction Evaluation

To deepen our interpretation of model performance, we conducted a regional analysis using the Yeo17 atlas with 100 cortical nodes [28, 29]. This atlas was registered, down-sampled, and reshaped to match the output brain states' size, using the same method we used to reshape the fMRI data with cropping and padding.

For each brain region, we calculated the regional MSE. To illustrate the model's performance across a range of error levels, we selected three representative regions: (1) the left parietal medial (ParMed, node 1), with an MSE close to the overall average; (2) the left superior parietal lobule (SPL, node 1), with below-average error; and (3) the right orbitofrontal cortex (OFC, node 1), with above-average error. For each region, we extracted the time series of one test sample and visualized the input, predicted, and ground truth brain activity. This analysis provided insight into how the model captures temporal dynamics at the regional level and highlighted variation in prediction performance across functionally distinct brain areas.

## III. RESULTS

### A. Model Convergence

During training, we monitored three key metrics: combined loss, masked MSE, and SSIM loss. As shown in Fig. 2, all metrics exhibited a substantial decline within the first eight epoch, indicating rapid initial learning. After approximately 20 epochs, the metrics stabilized, suggesting convergence. The final model was trained for 24 epochs.

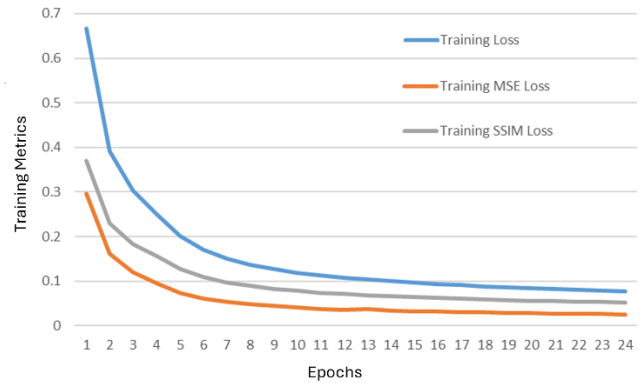

Fig. 2. This figure shows the changes of loss, masked MSE, and SSIM during the training process on training dataset. The blue, orange and green curves show the combined loss, masked MSE, and scaled SSIM loss, respectively.

### B. Quantitative Prediction Metrics

Using 32 consecutive brain states as input, our model successfully predicted the next 10 brain states with an overall masked MSE of 0.035 on the held-out test set. The corresponding unscaled SSIM loss was 0.046, indicating a high structural similarity between the predicted and true volumes, with the equivalent SSIM score of 0.954, close to the idea value of 1. All metrics were averaged across all test samples and all predicted time points.

Shuffling input brain states made the MSE rise dramatically to 4890 highlighting the critical role of temporal dependencies in brain state prediction.

To ensure the model was not merely replicating prior brain states, we computed the shifted MSE, the error between the predicted brain state and its previous brain state. The shifted MSE was around 0.076, more than twice the masked prediction MSE, confirming that the model was generating new brain states rather than repeating the past ones.

We further examined prediction accuracy across each of the 10 output brain states from $t_{T+1}$ to $t_{T+10}$. Fig. 3A shows the masked MSE was lowest for the first five predicted brain states (all below 0.01), but gradually increased in prediction time horizon, reaching a maximum of 0.144 at the $10^{th}$ brain state. In addition, the spread of the boxplots increased over time, particularly in the last five time points, which is consistent with the accumulation of prediction errors.

Furthermore, for each time step, we assessed spatial distribution of voxel-wise prediction errors by visualizing the absolute difference between the predictions and the corresponding real brain states as shown in Fig. 3B. No obvious spatial bias was observed. The temporal patterns of voxel level error also aligned with the MSE trend, showing minimal errors for the first five to six predicted brain states and increasing errors for later time points.

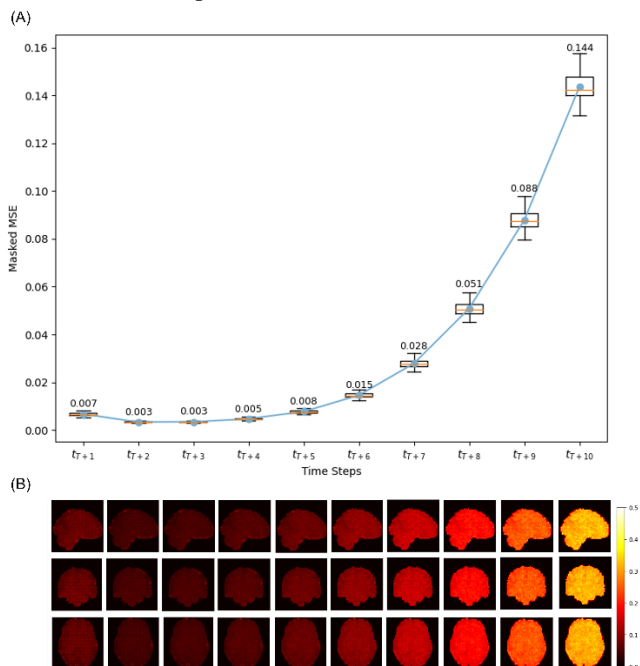

Fig. 3. Prediction errors across time. (A) Masked mean squared error (MSE) across test samples for each of the 10 predicted consecutive time points (from $t_{T+1}$ to $t_{T+10}$). Boxplots represent the distribution of masked MSE values at each time point, and the blue curve denotes the MSE per time point. On top of each boxplot, we show the corresponding MSE value rounded to three decimal places. (B) Prediction error on the voxel level. This part shows the absolute error from $t_{T+1}$ to $t_{T+10}$ (from left to right corresponding to the x-axis in (A)) for one test sample from orthogonal anatomical planes. All subfigures used the same color scale from 0 to 0.5, with darker colors implying lower error and brighter colors showing higher errors.

### C. Predicted Brain States

Although the low MSE suggested accurate predictions, it is still possible for a model to achieve low error by predicting average intensity or noise that lack meaningful spatial structure. To ensure that the predicted brain states retained contrast patterns, we visualized slices of different anatomical planes from an example prediction. As shown in Fig. 4, the predicted brain states closely resemble the ground truth, with clear and well-preserved contrast at different brain locations. Moreover, predicted dynamic changes in contrast also resemble the ground truth. This qualitative assessment was done among many randomly selected predictions and the conclusion is consistent.

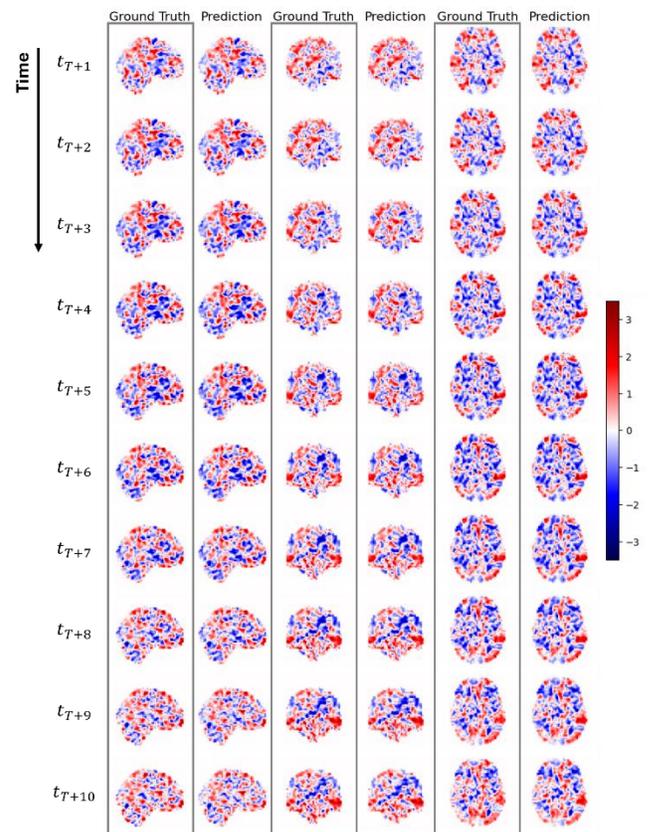

Fig. 4. Visualization of predicted and ground truth brain states of one example sample. Each row displays one of 10 consecutive time points predicted by the model. The left column (within gray boxes) of each pair shows the ground truth state at time $t_i$, while the right column shows the model's corresponding prediction. We display data for a sagittal, coronal and axial slice. The color scale is consistent across all subplots and centered at zero to highlight deviations in both directions. This figure demonstrates the model's ability to reconstruct temporally continuous brain states from fMRI.

### D. Regional Analysis

For each brain region included in the Yeo17 atlas (with 100 regions), we calculated the average MSE across all test samples. These values are visualized in Fig. 5A, where regions with lower prediction errors are shown in darker red and those with higher errors in brighter yellow. The distribution of all regional MSE was summarized into a boxplot shown in Fig. 5B, showing no outliers. The lowest regional MSE was observed in the left SPL (MSE = 0.028), while the highest occurred in the right OFC (MSE = 0.040).

To further investigate regional patterns, we ranked all regions by their average MSE and identified the top 10 regions with the highest and lowest prediction accuracies (Fig. 5C and 5D, respectively; full details in Table I). Regions with the most accurate predictions (Fig. 5D) were primarily located within the dorsal attention network (DorsAttn) and control network (Cont), with some regions from the salience/ventral attention network (SalVentAttn) and default mode (Default) network. Notably, these included SPL, inferior parietal lobule (IPL), post central (PostC), intraparietal sulcus (IPS), and lateral prefrontal cortex (PFCl).

In contrast, regions with the highest errors (Fig. 5C) are mainly presented in limbic regions, particularly the OFC and temporal pole (TempPole). Other high-error regions included the parahippocampal cortex (PHC) in Default network, insula (Ins) in SalVentAttn, S2 in somatomotor network (SomMot), and temporal-parietal junction (TempPar).

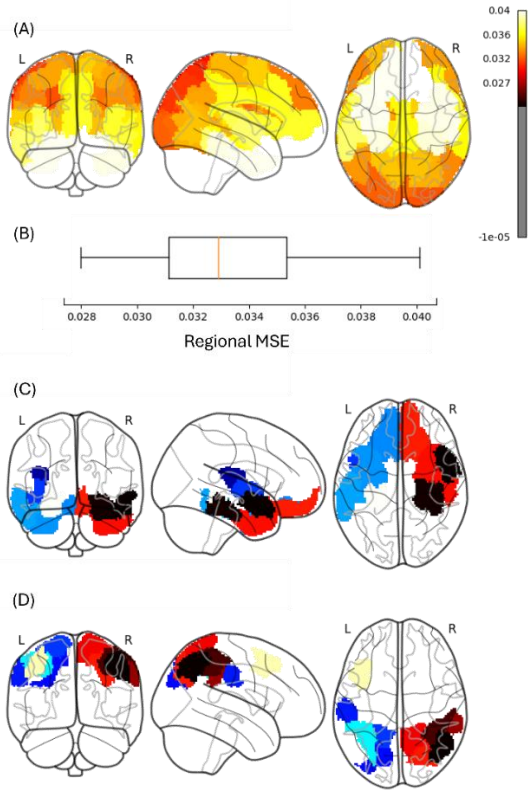

Fig. 5. Prediction MSE of brain regions. (A) shows the regional MSE using the Yeo17 atlas. Darker red colors imply a lower MSE, i.e. more accurate predictions. Brighter yellow colors stand for higher MSE, i.e. less accurate predictions. (B) Boxplot of MSE of 100 regions in the Yeo17 atlas. (C) Color-coded 20 brain regions with the highest MSE. (D) Color-coded 20 brain regions with the lowest MSE. For (C) and (D), colors are not assigned randomly to show different regions in the atlas.

TABLE I
TOP 10 REGIONS WITH THE LOWEST MSE (RIGHT) AND TOP 10 REGIONS WITH THE HIGHEST MSE (LEFT) IN THE YEO17 ATLAS

| Hemisphere | Network | Node | MSE | Hemisphere | Network | Node | MSE |
|---|---|---|---|---|---|---|---|
| R | Limbic B | OFC_1 | 0.0401 | L | DorsAttn A | SPL_1 | 0.0280 |
| L | Limbic B | OFC_1 | 0.0399 | R | Cont B | IPL_1 | 0.0282 |
| L | Limbic A | TempPole_1 | 0.0398 | R | DorsAttn A | SPL_1 | 0.0286 |
| R | Limbic A | TempPole_1 | 0.0396 | L | DorsAttn B | PostC_3 | 0.0287 |
| L | Default C | PHC_1 | 0.0395 | L | DorsAttn B | PostC_1 | 0.0292 |
| R | Default C | PHC_1 | 0.0394 | R | DorsAttn B | PostC_2 | 0.0296 |
| L | SalVentAttn A | Ins_1 | 0.0376 | L | Cont A | IPS_1 | 0.0296 |
| L | Limbic A | TempPole_2 | 0.0376 | R | SalVentAttn B | IPL_1 | 0.0297 |
| L | SomMot B | S2_1 | 0.0370 | R | Cont A | IPS_1 | 0.0298 |
| R | TempPar | TempPar_1 | 0.0367 | L | Default B | PFCl_1 | 0.0300 |

To further examine temporal prediction accuracy, we visualized the input, ground truth, and predicted brain dynamics as time series for three representative brain regions (Fig. 6). Among them, the left ParMed held a MSE of 0.035, close to the overall MSE. The other two regions, left SPL and right limbic OFC, had the MSE of 0.028 and 0.040, respectively. These time series further confirmed the model's ability to predict temporally coherent brain activity.

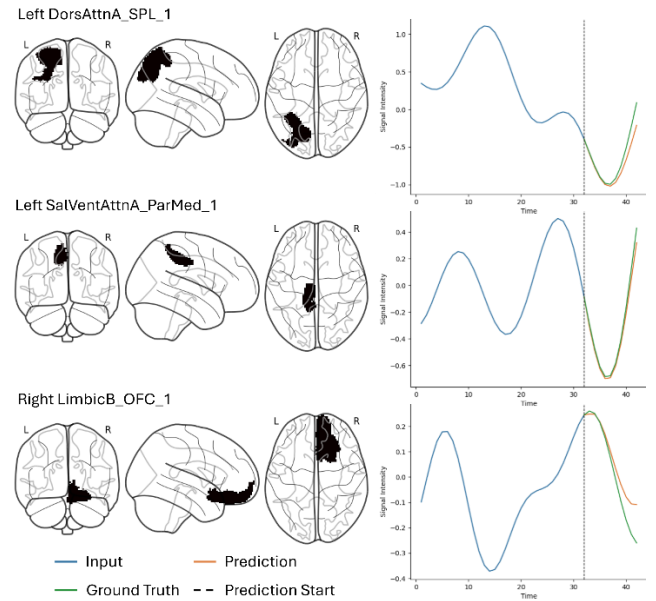

Fig. 6. Signal trajectories from three brain regions (black regions on glass brains on the left) showing predicted and ground truth fMRI signals along the time (on the right). Blue periods indicate the real fMRI signal as inputs, followed by green curves represent the ground truth, and orange curves represent the predicted brain states. The vertical dashed line at time-point 32 marks the onset of the prediction.

## IV. DISCUSSION

### A. Prediction Performance

Our model achieved high prediction accuracy (low overall masked MSE) at the whole-brain level, indicating that the predicted brain states are numerically close to the ground truth. Complementing this, the high SSIM score suggests that the model not only preserves intensity values but also retains important spatial structure and contrasts across brain regions. These two metrics demonstrate that the model generates voxel-wise predictions that are both numerically accurate and structurally plausible.

When we shuffle the input brain states, the prediction MSE dramatically increased by several orders of magnitude. This highlights that the model relies on temporal continuity and is not simply capturing the existence of different brain states in the input. In other words, our model learned meaningful temporal dependencies and utilized them to predict future brain states.

Furthermore, the shifted MSE, computed between the predicted states and the previous brain state, was substantially higher than the normal masked MSE. This suggests that the model is not merely replicating the adjacent volume but is instead generating new, temporally plausible brain states. The differences between the shifted and predicted MSE confirm the model's ability to forecast future dynamics rather than performing simple duplication.

In addition to the overall performance on the 10 predicted brain states, the prediction accuracy for individual brain states provides more insights. We observed the masked MSE gradually increases with the prediction time. This is actually a common challenge in autoregression tasks [30] and has also been found for region-based brain state prediction [21]. We also observed increasing variance in the prediction error across test samples for later time points, as indicated by the widening spread of the boxplots in Fig. 3A. This trend suggests growing uncertainty as the model forecasts further into the future.

Spatially, voxel-wise prediction error maps (Fig. 3B) showed no dominant bias, with errors distributed across the brain, indicating the spatial generalizability of the model across the whole brain.

### B. Predicted Brain State

Comparing the predicted brain states versus the ground truth shown in Fig. 4, our model preserves the spatial contrast patterns remarkably well. From the first time point ($t_{T+1}$) to the last time point ($t_{T+10}$), the gradual and consistent shifts in spatial intensity are clearly observable, indicating the model successfully captures the evolving spatial features of brain activity. The preservation of fine-grained dynamic contrast in 3D volume space suggests that the model has learned a robust internal representation of brain dynamics that generalizes beyond training samples. Although the MSE increases in later time points, the spatial similarity of predictions remains high. This suggests that there are some localized mismatches instead of the breakdown of spatial coherence.

### C. Regional Variation in Prediction Accuracy

Our regional level analysis revealed notable variation in prediction accuracy across different brain areas. The lowest regional MSE was observed in the left SPL (MSE = 0.028), while the highest error was almost doubled the lowest MSE and was found in right OFC (MSE=0.040). Even though they showed a gap, the prediction performance is still relatively stable across regions as there is no outlier found in Fig. 5B.

Based on the regional ranking results, the top 10 regions with the best predictions primarily belonged to DorsAttn and Cont networks, including regions like SPL, PostC, IPL, and IPS. These regions are known to support attention [31-33], visuospatial integration [31], spatial awareness [32], and proprioceptive processing [34]. Although no explicit task is performed during rest, individuals continue to receive sensory information (e.g. tactile input, temperature). Thus, these relevant brain regions may maintain structured and meaningful intrinsic patterns in the BOLD signals rather than random noise. Such patterns may be easier for the model to learn and predict over time. Additionally, the resting-state functional connectivity of these regions with other brain networks or regions, as observed in rs-fMRI studies [35-38], may give rise to structured dynamics that enhance predictability, allowing the model to more easily learn and generalize from regular spatiotemporal patterns compared to regions with more variable or loosely defined connectivity.

We also identified the PFCl in Default network as one of the regions with strong prediction performance. PFCl is responsible for high level functions, such as sensorimotor integration, attention, and memory [39]. A prior study found the amplitude of low-frequency fluctuations in dorsal PFCl is reliable at resting state [40], suggesting stable intrinsic dynamics that are favorable for temporal prediction modeling.

In contrast, regions with the highest prediction MSE were largely associated with the limbic network, particularly OFC and TempPole, which are known for their involvement in emotion regulation, reward processing, face recognition and social behavior [41, 42], functions that may not be engaged at rest. Consequently, their resting-state activities may contain more variability and larger inter-subject differences, making it difficult for the model to identify generalizable patterns.

Similarly, PHC, associated with visuospatial processing and episodic memory [43], showed elevated MSE. This may be explained by its high temporal variability at resting state [44]. Other high-error regions included Ins, S2 in SomMot, and TempPar, related to emotion, sensory, and attention [45-47]. Despite similar functions to some low-error regions, the increased complexity, dynamic integration, and individual variability in these areas likely introduce varying patterns that are difficult for the model to capture.

Moreover, OFC, anterior medial temporal lobe, and basal anterior temporal lobe are vulnerable to signal loss and susceptible artifacts as reported in [48-50], which further compromise the reliability of their BOLD signals and reduce prediction accuracy. Collectively, these challenges, including functional variability and signal quality, likely contribute to the relatively worse performance of the model in these regions.

Regarding the predicted brain states in time domain as time series (Fig. 6), the predicted brain states in the time domain closely tracked the overall trend of the real fMRI time series across all the selected regions. Notably, the first half of the predicted time series exhibited excellent correspondence with the actual brain activity, indicating that the model was able to effectively capture and extrapolate short-term temporal dynamics. However, as we predict longer time, a gradual divergence between the predicted and actual signals emerged, particularly in the later time points. This temporal degradation in accuracy is consistent with the trend observed in the MSE of individual brain state (Fig. 3). The growing discrepancy reflects the model's uncertainty accumulation over longer forecasts. Nevertheless, the smoothness and preserved fluctuation patterns of the predicted time series suggest that the model maintains the good dynamics in brain states rather than producing noisy or unstable outputs.

In general, while our model exhibited variations in performance across brain regions, likely due to different functional roles and signal properties, it demonstrated robust and consistent predictive accuracy across the whole brain.

## V. CONCLUSION

In this work, we proposed a novel Swin transformer based architecture that takes advantages from both SwiFT and convolutional decoders to predict brain states, defined as the voxel-level fMRI signals. Our model demonstrated fine-grained and accurate prediction, highlighting the potential of transformer-based models to effectively capture spatiotemporal dynamics in rs-fMRI data.

In the future, we aim to extend this work toward more accurate and longer-term predictions, which remains a key challenge in neural time series modelling. Beyond predictive performance, interpreting the learned temporal dependencies may offer new insights into intrinsic brain dynamics and organizations, with the potential in reducing fMRI scan time, improving our understanding of brain state evolvement and transitions, improving our understanding of brain disorders, and contributing to the advancement of BCI.